# Spin Relaxation in Single Layer Graphene with Tunable Mobility

Wei Han, Jen-Ru Chen, Deqi Wang, Kathleen M. McCreary, Hua Wen, Adrian G. Swartz, Jing Shi, Roland K. Kawakami[†]

*Department of Physics and Astronomy, University of California, Riverside, CA 92521*

[†]e-mail: roland.kawakami@ucr.edu

**Abstract:**

Graphene is an attractive material for spintronics due to theoretical predictions of long spin lifetimes arising from low spin-orbit and hyperfine couplings. In experiments, however, spin lifetimes in single layer graphene (SLG) measured via Hanle effects are much shorter than expected theoretically. Thus, the origin of spin relaxation in SLG is a major issue for graphene spintronics. Despite extensive theoretical and experimental work addressing this question, there is still little clarity on the microscopic origin of spin relaxation. By using organic ligand-bound nanoparticles as charge reservoirs to tune mobility between 2700 and 12000 cm$^2$/Vs, we successfully isolate the effect of charged impurity scattering on spin relaxation in SLG. Our results demonstrate that while charged impurities can greatly affect mobility, the spin lifetimes are not affected by charged impurity scattering.

Single layer graphene (SLG) is a promising material for spintronics due to theoretical predictions of long spin lifetimes based on its low intrinsic spin-orbit and hyperfine couplings[1-5]. However, spin lifetimes measured in SLG spin valves are much shorter (0.05 – 1.2 ns)[6-9] than predicted (100 ns – 1 μs)[1-5]. Thus, the origin of spin relaxation in SLG has become a central issue for graphene spintronics and has motivated intense theoretical and experimental studies. Theoretical studies of spin relaxation include impurity scattering[10], ripples[5], spin orbit domains[11,12], and substrate effects[13], while experimental studies have investigated contact-induced spin relaxation[7,9,14], ripples[15], band structure effects[6,14,16], edge effects[7] and charged impurity scattering[6,8]. However, apart from recognizing the requirement for high quality tunneling contacts to suppress contact-induced spin relaxation[9], there is little clarity regarding the origin of spin relaxation. To address the situation, it is crucial to develop experimental techniques that systematically isolate the various microscopic sources of spin relaxation.

In this work, we successfully isolate the effect of charged impurity scattering on spin relaxation in SLG by exploiting the novel tunable mobility imparted by organic ligand-bound nanoparticles on the SLG surface[17]. The nanoparticles act as charge reservoirs that freely transfer charge with graphene at room temperature. At low temperature, the frozen charge distribution on the nanoparticles results in SLG mobility ranging from 2700 to 12000 cm$^2$/Vs. This approach is able to isolate the effect of charged impurity scattering on spin relaxation more clearly than previous investigations based on adatom deposition[8]. This is because depositing adatoms to the graphene surface could introduce additional effects such as short-range scattering, lattice deformation, and/or spin-orbit coupling, whereas such effects should be minimized in the current approach. Additionally, we utilize tunnel barriers to suppress contact-induced effects in order to investigate spin relaxation with high sensitivity. At fixed carrier concentration, our results show that spin lifetime exhibits little variation as mobility is tuned between 2700 and 12000 cm$^2$/Vs. This demonstrates that while charged impurities can greatly affect mobility, spin is very robust against this type of scattering. Specifically, for spin lifetimes below 2 ns, charged impurity scattering does not induce either Elliot-Yafet (EY) or Dyakonov-Perel (DP) spin relaxation at a level that affects spin lifetime.

Experiments are performed on SLG spin valves with device geometry illustrated in Figure 1a. Graphene flakes are mechanically exfoliated from highly ordered pyrolytic graphite (SPI supplies, ZYA grade) onto an SiO$_2$ (300 nm thickness)/Si substrate[18]. SLG is identified by optical microscopy and Raman spectroscopy[19]. First, two Au electrodes are fabricated on the two ends of SLG using e-beam lithography. Then, ferromagnetic (FM) Co electrodes with tunnel barriers (MgO/TiO$_2$) are fabricated using a second step of e-beam lithography and angle evaporation (electrical characterization of the tunneling contact is shown in supplementary information). Typically several Co electrodes are fabricated with widths between 80 nm and 300 nm to have different coercivities, but only two Co electrodes are wired up for the spin transport measurements. Details of device fabrication are provided elsewhere[14, 20]. After fabrication, SLG spin valves are decorated with organic ligand-bound nanoparticles, which will be used as a charge reservoir to tune the mobility of SLG[17]. The nanoparticles are iron oxide, mostly γ-Fe$_2$O$_3$ with a diameter of 13 nm and coated with the organic ligand (oleic acid). To control the density of nanoparticles on graphene, they are diluted in toluene with volume concentrations ranging from 1:10 to 1:2000, similar to previous studies[17]. After exposing the device to several drops of nanoparticle solution, the sample is maintained at room temperature in ambient conditions until it is completely dried.

Electrical measurements of SLG spin valves are performed using AC lock-in techniques in an Oxford He-4 system with variable temperature (2-300 K) and magnetic field (0-7 T) capabilities. Electrical properties are characterized by measuring resistivity at different carrier concentrations, controlled by the application of backgate voltage (Vg). During the measurement, an AC current of 1 μA (13 Hz) is applied between the two Au electrodes, while the voltage is measured between the two central Co electrodes. Due to the highly resistive nature of the molecular links, addition of nanoparticles on SLG does not provide a parallel current path. The nanoparticles act as a charge reservoir that can donate or accept electrons,[17] and at room temperature the charge state adjusts with a time scale of several minutes. This behavior is distinct from adatom dopants that typically provide a fixed electronic doping. As $V_g$ is swept from –50 V to +50 V at a rate of ~0.2V/s (black curve in Fig. 1b), the resistivity peak (i.e. Dirac point, $V_D$) is located at $V_g$ = – 28 V. On the contrary, when $V_g$ is swept from +50 V to –50 V with the same rate (red curve in Fig. 1b), the resistivity peak is located at $V_g$ = - 10 V, which indicates that the nanoparticles are acceptors. Hence, the observed dependence of Dirac point on voltage ramp direction demonstrates that the gate voltage alters the charge state in the nanoparticles, which in turn results in hysteresis in the SLG resistivity. In general, we observe that the Dirac point voltage gradually shifts to become equal to the applied gate voltage. Although the microscopic reason for this behavior is not fully understood, it nevertheless indicates that the charge distribution of the nanoparticles gradually adjusts to bring the graphene toward charge neutrality. In addition to neutralizing the charge induced by the backgate, the charge transfer also neutralizes (to the best of its ability) electron-hole puddles from charged impurities in the $SiO_2$ substrate and surface residues[21, 22]. As a result, the mobility of the SLG can be tuned. A high mobility is achieved by applying a gate voltage (e.g. -10 V for "A" state in Figure 1b) and waiting for ~1 hour to let the system reach steady state, followed by a rapid cooling to 10 K (~30 min) to freeze in the optimized charge distribution. Figure 1c shows the gate-dependent resistivity at low temperature following the cooling procedure described above. At 10 K, the up and down sweeps only show one resistivity peak at $V_g$ = -10 V, indicating that the charge transfer is halted at low temperatures. The mobility (the average of the electron and hole mobilities) is determined based on the linear dependence of conductivity on gate voltage away from the Dirac point[23]. To determine the SLG mobility ($\mu$), we first calculated the carrier concentration, $n$ (positive for holes), which is directly related to $V_g$ by $n = -\alpha(V_g - V_D)$ with $\alpha = 7.2 \times 10^{10} \text{V}^{-1}\text{cm}^{-2}$. Then, we plot the SLG conductivity as a function of the carrier concentration, as shown in Fig. 2a. The SLG mobility is calculated to be 7000 cm$^2$/Vs, using the average of the electron and hole mobilities ($\mu = (\mu_e + \mu_h)/2$), which are determined by linear fitting the regime close to the Dirac point ($\mu_{e,h} = |\Delta\sigma/e\Delta n|$). The SLG mobility for data in Figure 1c and 2a is calculated to be 7000 cm$^2$/Vs. The conductivity exhibits electron-hole asymmetry, which is discussed further below.

Lower mobilities are obtained by altering the cooling procedure. Starting from a steady-state charge distribution at $V_g$ = -10 V ("A" state), the gate voltage is changed to a different value and the sample is quickly cooled to 10 K while holding $V_g$ constant. Because the nanoparticle charge distribution is not allowed to reach steady state, it will not be optimized and therefore produce a lower mobility. Figures 2a-2d show SLG mobility of 7000, 5500, 4400, and 2700 cm$^2$/Vs, respectively, resulting from different pre-cooling procedures. The highest mobility (7000 cm$^2$/Vs) is achieved by cooling down directly from the "A" state at room temperature ($V_g$ held at -10 V). For the lowest mobility (2700 cm$^2$/Vs), the sample is stabilized in the "A" state and then quickly cooled down after switching the gate voltage to $V_g$ = +50 V. By systematically varying the charged impurity scattering in a single device, this provides a unique way of studying the effects of charged impurity scattering on spin relaxation in SLG.

The electron-hole asymmetry in Figure 2a may be related to resonant scattering induced by the nanoparticles (on a test flake, the Raman D peak increases slightly from $I_D/I_G$ = 0.05 to 0.08 when nanoparticles are introduced), with asymmetry decreasing as charged impurity scattering becomes more dominant (Fig. 2b-2d). Possible resonant scattering, however, is not significant for spin relaxation, as spin lifetimes do not change with doping by nanoparticles (Figure 5).

Studies of the spin relaxation are performed on SLG spin valves consisting of two spin-sensitive Co electrodes (E2, E3) and two Au electrodes (E1, E4) via nonlocal Hanle spin precession measurements (Figure 3a, right). Nonlocal voltages ($V_{NL}$) are measured between E3 and E4 using lock-in detection with an AC injection current of $I$ = 1 µA rms at 13 Hz applying across E2 and E1. Prior to the Hanle measurement, nonlocal magnetoresistance (MR) measurement is performed (device A before doping) with the magnetic field in plane to achieve parallel and antiparallel alignments of the Co electrodes (Figure 3a, left). Nonlocal MR scans are shown in Figure 3b. Then, an out-of-plane magnetic field ($B_\perp$) is applied, which causes the spins to precess as they diffuse from E2 to E3 (Fig. 3a, right). Fig. 3c shows the characteristic Hanle curves at 300 K at the Dirac point, in which the red circles (black circles) are for the parallel (antiparallel) alignment of the Co magnetizations. Taking into account that the contact width of the Co electrodes is only ~50 nm, which is much shorter compared to the spin diffusion length, we approximate the electrode as a point contact and use the following equation[24] to fit the Hanle curve and determine the spin lifetime ($\tau_s$), diffusion coefficient ($D$), and spin diffusion length ($\lambda_s = \sqrt{D\tau_s}$):

$$R_{NL} \propto \pm \int_0^\infty \frac{1}{\sqrt{4\pi Dt}} \exp\left[-\frac{L^2}{4Dt}\right] \cos(\omega_L t) \exp(-t/\tau_s) dt \quad (1)$$

where the + (-) sign is for the parallel (antiparallel) magnetization state. $\varpi_L = g\mu_B B_\perp / \hbar$ is the Larmor frequency, $g = 2$ is the electron g-factor, $\mu_B$ is the Bohr magneton, and $\hbar$ is the reduced Planck's constant. The best fit (solid lines in Fig. 3c) yields $D = 1.0 \times 10^{-2}$ m$^2$s$^{-1}$ and $\tau_s = 559$ ps, which correspond to a spin diffusion length of $\lambda_s = \sqrt{D\tau_s} = 2.4$ μm.

To systematically study the spin dependent properties with SLG of different mobilities, we perform Hanle spin precession on the same SLG spin valve (device A) decorated by a layer of ligand-bound nanoparticles. For mobilities of 7000, 5500, 4400 and 2700 cm$^2$/Vs, the spin lifetimes and diffusion coefficients obtained via Hanle spin precession are summarized in Fig. 4a and 4b. First, a similarity is observed that for all the different mobilities, the spin lifetimes exhibit a minimum (0.5 ns) at the Dirac point, and longer spin lifetimes up to 1.8 ns are observed at high carrier (electron or hole) densities. More interestingly, the spin lifetimes do not increase even though the mobility changes by a factor of 2.6. On the contrary, the diffusion coefficients are relatively larger for the SLG spin valve with higher mobility, as shown in Fig. 4b. To directly compare momentum scattering and spin scattering, we estimate the momentum scattering time ($\tau_p$) within Boltzmann transport theory[25] in the regime of carrier concentrations from $1 \times 10^{12}$ cm$^{-2}$ to $3.6 \times 10^{12}$ cm$^{-2}$:

$$\tau_p(n) = \frac{h\sigma}{e^2 v_F \sqrt{n g_s g_v \pi}} \quad (2)$$

where $h$ is Planck's constant, $\sigma$ is the conductivity of SLG, $e$ is the electron charge, $v_F$ is the Fermi velocity (~ $10^6$ m/s), and $g_v$ and $g_s$ are valley and spin degeneracies. For the different mobilities, the relationship between $\tau_s$ and $\tau_p$ at fixed carrier concentration is plotted in Figure 4c (electron doping at 2.16 × $10^{12}$ cm$^{-2}$ and 3.60 × $10^{12}$ cm$^{-2}$) and Figure 4d (hole doping at 2.16 × $10^{12}$ cm$^{-2}$ and 3.60 × $10^{12}$ cm$^{-2}$). It is clearly shown that $\tau_s$ exhibits little variation while $\tau_p$ changes by as much as a factor of three. The different behaviors of spin transport properties (spin lifetimes) and charge transport properties (mobility, diffusion coefficient and momentum scattering time) indicate that different factors are important for spin and charge transport. Specifically, while charged impurity scattering has a strong effect on mobility, spin is very robust against this type of scattering.

We further investigate spin relaxation in devices with higher mobility (device B). Figure 5 shows the gate-dependent spin lifetime for mobility tuned between 4200 and 12000 cm$^2$/Vs and the data for pristine graphene spin valve with mobility of 4050 cm$^2$/Vs. Again, the measured spin lifetimes lie in the same range of 0.5-1.8 ns for both the 4000 cm$^2$/Vs and 12000 cm$^2$/Vs cases. We also compare this with the gate-dependent spin lifetime before depositing the nanoparticles onto the graphene (Fig. 5, open squares, dashed line). These pristine graphene samples have similar spin lifetimes and gate dependence, which indicate the nanoparticles themselves do not introduce any substantial spin relaxation. Therefore, from the point of view of spin relaxation, the nanoparticle-doped graphene is representative of pristine graphene spin valves. These results support that charged impurity scattering is not the dominant source of spin relaxation in SLG. We note that this conclusion is valid for spin lifetimes in the ~1 ns range; for example, if spin valves with much longer lifetimes are developed, it may turn out that charged impurity scattering becomes significant at the longer time scale.

Our main result that $\tau_s$ exhibits little variation as $\tau_p$ is tuned by charged impurity scattering (Fig. 4c, 4d) is consistent with previous investigations showing that $\tau_s$ varies linearly with $D$ (~$\tau_p$) as a function of carrier concentration[6, 14]. This linear relation has been taken as evidence that spin relaxation in SLG originates from an Elliot-Yafet-like (EY) process (i.e. finite probability of spin-flip during a momentum scattering event). The reason our current result is consistent with the previous work is that EY-like spin relaxation could be generated by various microscopic mechanisms, such as short-range scattering, resonant scattering, charged impurity scattering, phonon scattering, and inhomogeneous Rashba field[11, 26]. If any of these mechanisms other than charged impurity scattering are responsible for the EY-like spin relaxation, then $\tau_s$ should scale with $D$ as carrier concentration is varied, while $\tau_s$ at constant carrier concentration remains unchanged as mobility is tuned by charged impurity scattering (Fig. 4c, 4d). Thus, the reason for the different dependencies of $\tau_s$ is that the mobility tuning focuses on the effect of charged impurity scattering, while the carrier concentration tuning can include a broader set of microscopic mechanisms.

While these results provide support for EY-like spin relaxation over Dyakonov-Perel-like (DP) spin relaxation (i.e. spin precession between momentum scattering events), recent theoretical and experimental studies highlight the complexity of spin relaxation in graphene. Theoretical studies show that inhomogeneous spin-orbit coupling can produce both DP-like and EY-like scaling[11, 26], the relation between $D$ and $\tau_p$ contains explicit density dependence due to tunable Fermi energy[27], and $sp_3$ bonding can generate substantial spin relaxation[10]. Recent experiments on few layer graphene also find a variety of behaviors, with some groups reporting EY-like behavior[28] and other groups reporting DP-like behavior[14, 29]. Therefore, it becomes increasingly important to develop systematic experimental methods to isolate microscopic mechanisms that could contribute to spin relaxation.

In summary, spin relaxation in SLG due to charged impurity scattering is investigated using ligand-bound nanoparticles to tune mobility without altering the sample structure or adding dopants. No significant variation in spin lifetime is observed at constant carrier density as mobility is tuned between 2700 to 12000 cm$^2$/Vs. Our results demonstrate that spin is SLG is robust to charged impurity scattering and point out future directions to investigate spin relaxation in graphene.

**ACKNOWLEDGEMENT**

We acknowledge technical assistance from Yadong Yin. WH, JRC, KM, HW, AS, and RKK acknowledge the support of ONR (N00014-09-1-0117), NSF (DMR-1007057), and NSF (MRSEC DMR-0820414). DW and JS acknowledge the support of DOE award (DE-FG02-07ER46351).

**Figure Captions:**

Fig. 1: Charge transport properties of SLG spin valves (device A). (a) Schematic device geometry for a graphene spin valve. (b) Resistivity as a function of the gate voltage at room temperature. Two Dirac points indicate slow charge transfer between SLG and iron oxide nanoparticles. The black (red) curve shows the resistivity curve as the gate voltage is swept up (down). (c) Resistivity as a function of the gate voltage at 10 K, cooled down quickly from initial state "A" in Fig. 1b.

Fig. 2: Tuning of mobility for the same SLG spin valve (device A). (a-d) SLG conductivity as a function of carrier concentration at 10 K with mobility 7000 cm$^2$/Vs, 2700 cm$^2$/Vs, 5500 cm$^2$/Vs, and 4400 cm$^2$/Vs, respectively. The carrier concentration is calculated from the gate voltage using the relation by $n = -\alpha(V_g - V_D)$ with $\alpha = 7.2 \times 10^{10} \text{V}^{-1}\text{cm}^{-2}$. These different mobilities are obtained using different cooling procedures as described in the text.

Fig. 3: Hanle spin precession measurements for a SLG spin valve (device A). (a) Nonlocal MR (left) and Hanle measurement geometry (right). For nonlocal MR measurements, the magnetic field is applied in-plane along the easy axis of the Co electrodes. For Hanle measurements, the magnetic field is applied perpendicular to the SLG surface. (b) Representative nonlocal measurements of device A at 300 K. (c) Representative Hanle measurements of device A at 300 K. The red (black) circles are data taken for parallel (antiparallel) Co magnetizations. Solid lines are the best fit based on equation 1.

Fig. 4: Spin dependent properties of a SLG spin valve at 10 K (device A). (a) Spin lifetimes of device A with different mobilities varying from 2700 to 7000 cm$^2$/Vs as a function of carrier concentration. (b) Diffusion coefficient of device A with different mobilities varying from 2700 to 7000 cm$^2$/Vs as a function of carrier concentration. (c-d) The relationship of spin lifetime and momentum scattering time for electron and hole doping, respectively.

Fig. 5: Spin lifetime for device B as a function of carrier concentration for mobility of 4200 cm$^2$/Vs (black), 12000 cm$^2$/Vs (red) and pristine graphene with mobility 4050 cm$^2$/Vs (open black square). Measurements are performed at 10 K.

Figure 1

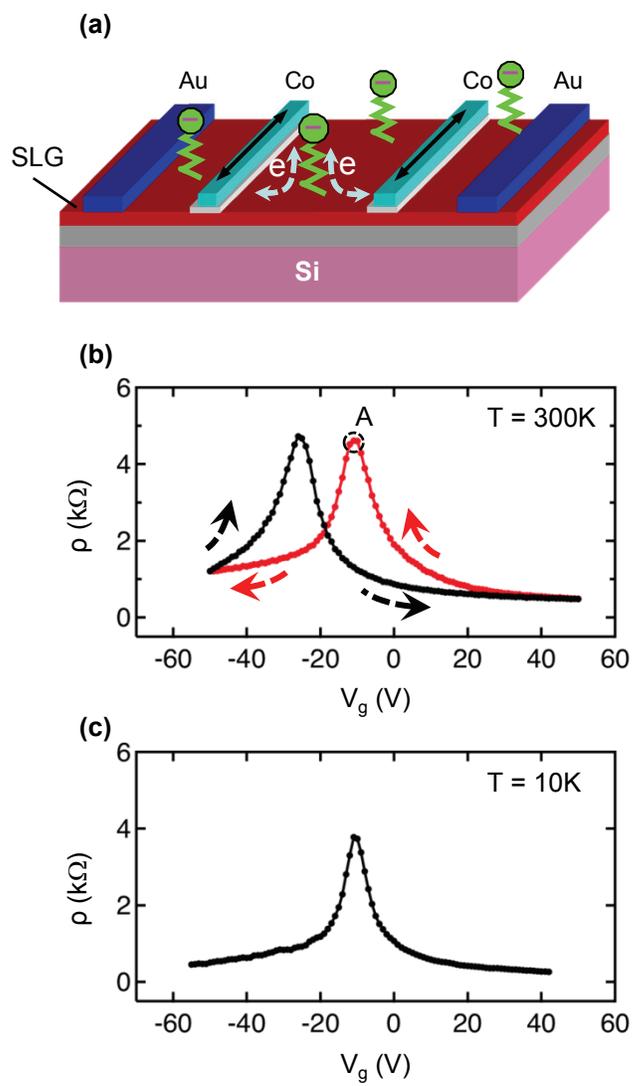

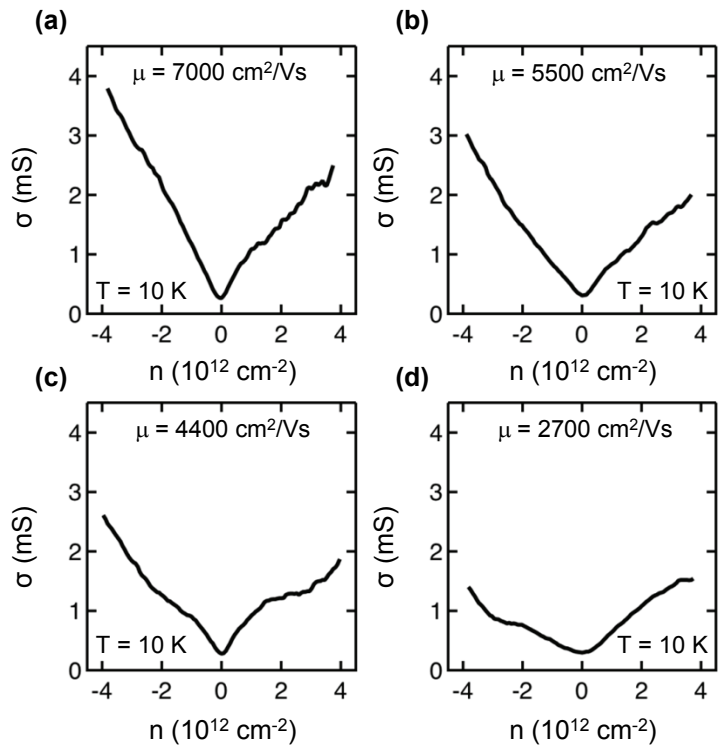

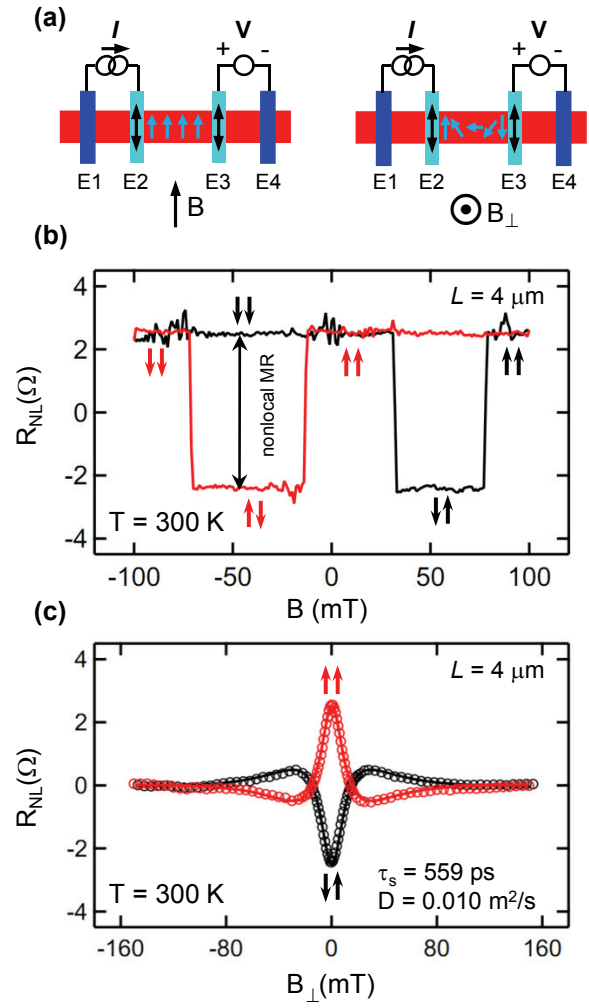

Figure 4

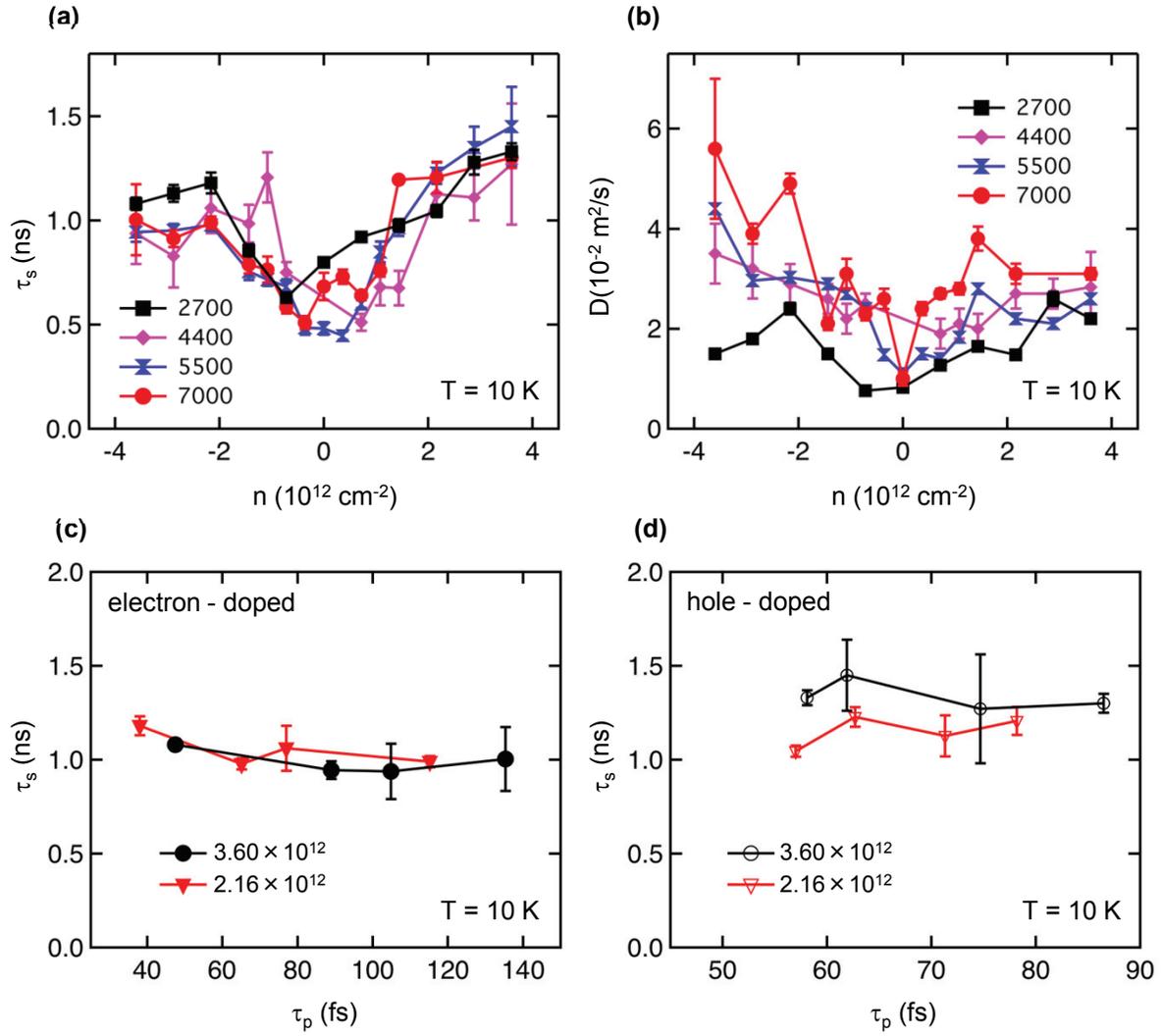

Figure 5

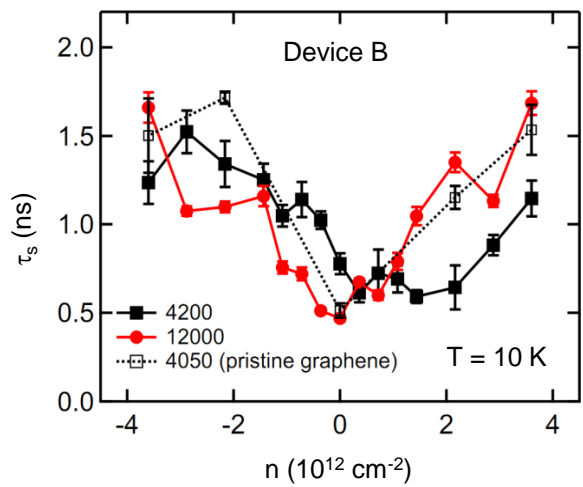